\documentclass[namedreferences]{solarphysics}
\usepackage[optionalrh]{spr-sola-addons} % For Solar Physics 
\usepackage{graphicx}        % For eps figures, newer & more powerfull
\usepackage{amssymb}        % useful mathematical symbols
\usepackage{color}           % For color text: \color command
\usepackage{url}             % For breaking URLs easily trough lines
\usepackage[pdfborder={0 0 0 },urlcolor=blue,breaklinks]{hyperref}
\ifx \doiurl    \undefined \def \doiurl#1{\href{http://dx.doi.org/#1}{\textsf{#1}}}\fi
\ifx \adsurl    \undefined \def \adsurl#1{\href{http://adsabs.harvard.edu/abs/#1}{\textsf{#1}}}\fi
\ifx \arxivurl  \undefined \def \arxivurl#1{\href{http://arxiv.org/abs/#1}{\textsf{#1}}}

\usepackage[varg]{txfonts}

\newcommand{\etal}{{\it et al.}}

\begin{document}
\begin{article}

\begin{opening}
\title{Complex Network Approach to the Statistical Features of the Sunspot
Series} \author{Yong Zou$^{1, 2, 3}$ \sep Michael
Small$^{4}$ \sep Zonghua Liu$^{1}$ \sep J\"urgen
Kurths$^{3, 5, 6}$
}
\runningauthor{Y. Zou \etal}
\runningtitle{Complex Network Perspective of the Sunspot Series}
\institute{$^{1}$Department of Physics, East China Normal University, Shanghai,
China, email: \url{yzou@phy.ecnu.edu.cn}\\
$^{2}$Department of Electronic and Information Engineering, Hong
Kong Polytechnic University, Hong Kong \\
$^{3}$Potsdam Institute for Climate Impact
Research, Potsdam, Germany \\
$^{4}$School of Mathematics and Statistics, University of
Western Australia, Crawley, Australia \\
$^{5}$Department of Physics, Humboldt
University Berlin, Berlin, Germany \\ 
$^{6}$Institute for Complex Systems and
Mathematical Biology, University of Aberdeen, Aberdeen, United Kingdom
}

\date{Received 12 April 2013, Revised \today.}

% %%%%%%%%%%%%%%%%%%%%%%%%%%%%%%%%%%%%%%%%%%%%%%%%%% %% Abstract
\begin{abstract}
Complex network approaches have been recently developed as
an alternative framework to study the statistical features of time-series
data. We perform a visibility-graph analysis on both the daily and monthly
sunspot series. Based on the data, we propose two ways to construct the network:
one is from the original observable measurements and the other is from a
negative-inverse-transformed series. The degree distribution of the derived networks for the strong maxima has
clear non-Gaussian properties, while the degree distribution for minima is
bimodal. The long-term variation of the cycles is reflected by hubs in the
network which span relatively large time intervals. Based on standard network structural measures, we propose to
characterize the long-term correlations by waiting times between two
subsequent events. The persistence range of the solar cycles has been identified over
15\,--\,1000 days by a power-law regime with scaling exponent $\gamma = 2.04$ of
the occurrence time of the two subsequent and successive strong minima. In contrast, a 
persistent trend is not present in the maximal numbers, although maxima do have
significant deviations from an exponential form. Our results suggest some new
insights for evaluating existing models. The power-law regime suggested by the
waiting times does indicate that there are some level of predictable patterns in
the minima.

\end{abstract}
%%%%%%%%%%%%%%%%%%%%%%%%%%%%%%%%%%%%%%%%%%%%%%%%%%%
%% Keywords
%
\end{opening}

\section{Introduction}\label{S-Introduction} 
Solar-cycle prediction, {\emph {i.e.}} forecasting the amplitude and/or the
epoch of an upcoming maximum is of great importance as solar activity has a
fundamental impact on the medium-term weather conditions of the Earth,
especially with increasing concern over the various climate change scenarios. However,
predictions have been notoriously wayward in the past~\cite{Hathaway2010,PesnellReview2012}.
There are basically two classes of methods for solar cycle predictions:
empirical data-analysis-driven methods and methods based on dynamo models. Most
successful methods in this regard can give reasonably accurate predictions only
when a cycle is well advanced ({\emph{e.g.,}} three years after the minimum) or
with the guidance from its past~\cite{Kurths1990,Hathaway1994}. Hence, these
methods show very limited power in forecasting a cycle which has not yet started. The
theoretical reproduction of a sunspot series by most current models shows
convincingly the ``illustrative nature" of the existing
record~\cite{Petrovay2010}. However, they generally failed to predict the slow
start of the present Cycle 24~\cite{Solanki2011}. One reason cited for this is
the emergence of prolonged periods of extremely low activity.
The existence of these periods of low activity brings a big challenge for
solar-cycle prediction and reconstruction by the two classes of methods
described above, and hence prompted the development of special ways to evaluate
the appearance of these minima~\cite{Brajsa2009}.
Moreover, there is increasing interest in the minima since they are known to
provide insight for predicting the next maximum~\cite{Ramesh2012}.

Some earlier authors have both observed and made claims for the chaotic or
fractal features of the observed cycles, but the true origin of such features
has not yet been fully resolved. For instance, the Hurst exponent has been used
as a measure of the long-term memory in time
series~\cite{Mandelbrot1969,Ruzmaikin1994}\,--\,an index of long-range
dependence that can be often estimated by a rescaled range analysis.
The majority of Hurst exponents reported so far for the sunspot numbers are well
above $0.5$, indicating some level of predictability in the data. Nonethteless,
it is not clear whether such predictability is due to an underlying chaotic
mechanism or the presence of correlated changes due to the quasi-11-year
cycle~\cite{OliverPRE1998,PesnellReview2012}. It is the irregularity (including
the wide variations in both amplitudes and cycle lengths) that makes the
prediction of the next cycle maximum an interesting, challenging and, as yet,
unsolved issue. In contrast to the 11-year cycle {\em per se}, we concentrate on
the recently proposed hypothetical long-range memory mechanism on time scales
shorter than the quasi-periodic 11-year cycle~\cite{Rypdal_JGR2012}.

In this work, we provide a distinct perspective on the strong maximal activities
and quiescent minima by means of the so-called visibility graph analysis. Such
graphs (mathematica graphs, in the sense of networks) have recently emerged as
one alternative to describe various statistical properties of complex systems. In
addition to applying the standard method, we generalize the technique further
-- making it more suitable for studying the observational records of the solar
cycles.

\section{Data and Network Construction}\label{S-Data} 
Both the International Sunspot Number (ISN) and the sunspot area (SSA)
series~\cite{sidcDataBelgium} are used in this work, and we have obtained
consistent conclusions in either case. The length of the data sets are
summarized in Table \ref{tab:tspan}. We perform a visibility-graph analysis
using both monthly and daily sunspot series, which yields, respectively,
month-to-month and day-to-day correlation patterns of the sunspot activities.
Note that we depict the annual numbers \emph{only} for graphical visualization
and demonstration purposes (we use the annual numbers to demonstrate our method
--- the actual analysis is performed in daily and monthly data).
We discuss the results with the ISN (in
Figs.~\ref{sn_sa_data}, \ref{ts_deg_maxMin_cp}) in the main text and illustrate
the results for the SSA (in Figs.~\ref{sa_nasa_data},
\ref{ts_deg_maxMin_cpNASA}) with notes in the captions. Moreover, we compare
our findings based on observational records to the results obtained from data
produced by simulations from computational
models~\cite{BarnesModel1980,MininniPRL2000}.
\begin{table}
\caption{Temporal resolution and the length of the data sets. Values in
 parentheses are the number of points of each series. Note that the annual ISN
 is used
  \emph{only} for graphical visualization purposes and to provide a reference time interval for models.}
 \label{tab:tspan} 
\begin{tabular}{lcc}
\hline
        & ISN                                & SSA                                \\
\hline
Day     & \small{1 Jan 1849 -- 31 Oct 2011} & \small{1 May 1874 -- 31 Oct 2011}\\
        & \small{(59473)}					 & \small{(50222)}         		      \\
\hline
Mon.    & \small{Jan 1849 -- Oct 2011}        & \small{May 1874 -- Oct 2011} \\
        & \small{(1944)}                     & \small{(1650)}                     \\
\hline
Year    & \small{1700 -- 2010}                & \small{not used}              \\
        & \small{(311)}                      &                                    \\
\hline
\end{tabular}
 \end{table}

Recently a variety of methods have been proposed for studying time series from a
complex networks viewpoint, providing us with many new and distinct statistical
properties~\cite{Zhang2006,Xu2008,Marwan2009,Donner2009,Donner2010IJBC}. In
this work, we restrict ourselves to the concept of the visibility graph (VG), where
individual observations are considered as vertices and edges are introduced
whenever vertices are visible. More specifically, given a univariate time series
$x(t_i)_{i= 1, \ldots,N}$, we construct the 0\,--\,1 binary adjacency matrix
$A_{N \times N}$ of the network. The algorithm for deciding non-zero entries of
$A_{i,j}$ considers two time points $t_i$ and $t_j$ as being mutually connected
vertices of the associated VG if the following criterion
\begin{equation} \label{vis_cond}
	\frac{x(t_i)-x(t_k)}{t_k - t_i} > \frac{x(t_i) - x(t_j)}{t_j - t_i}
\end{equation}
is fulfilled for all time points $t_k$ with $t_i < t_k < t_j$~\cite{Lacasa2008}.
Therefore, the edges of the network take into account the temporal information
explicitly. By default, two consecutive observations are connected and the graph
forms a completely connected component without disjoint subgraphs. Furthermore,
the VG is known to be robust to noise and not affected by choice of algorithmic
parameters -- most other methods of constructing complex networks from time
series data are dependent on the choice of some parameters ({\emph{e.g.}} the
threshold $\varepsilon$ of recurrence networks, see more details
in~\cite{Donner2009}).
While the inclusion of these parameters makes these alternative
schemes more complicated, they do gain the power to reconstruct (with sufficient
data) the underlying dynamical system. For the current discussion we prefer the
simplicity of the visibility graph. The VG approach is particularly interesting for
certain stochastic processes where the statistical properties of the resulting
network can be directly related with the fractal properties of the time
series \cite{Lacasa2009,Elsner2009,Lacasa2012,Donner2012,Donges2013}.

\begin{figure}
  \centering
   \includegraphics[width=\columnwidth]{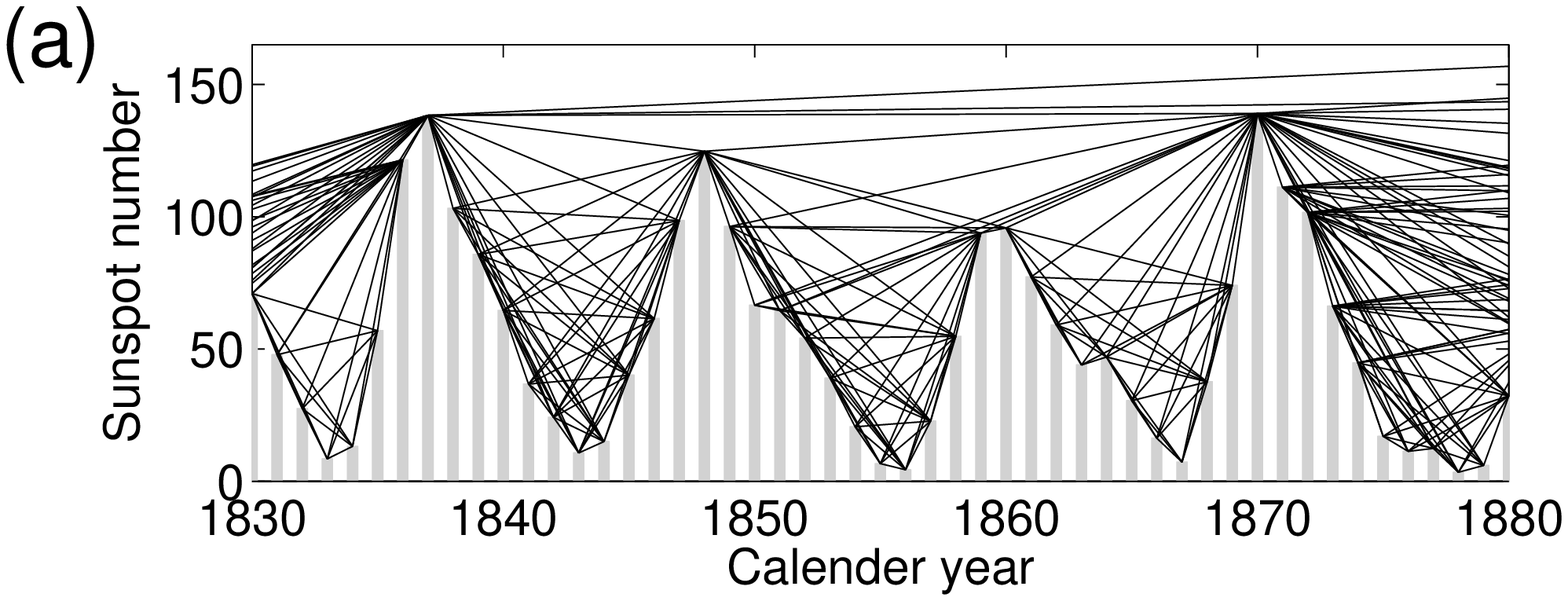}
   \includegraphics[width=\columnwidth]{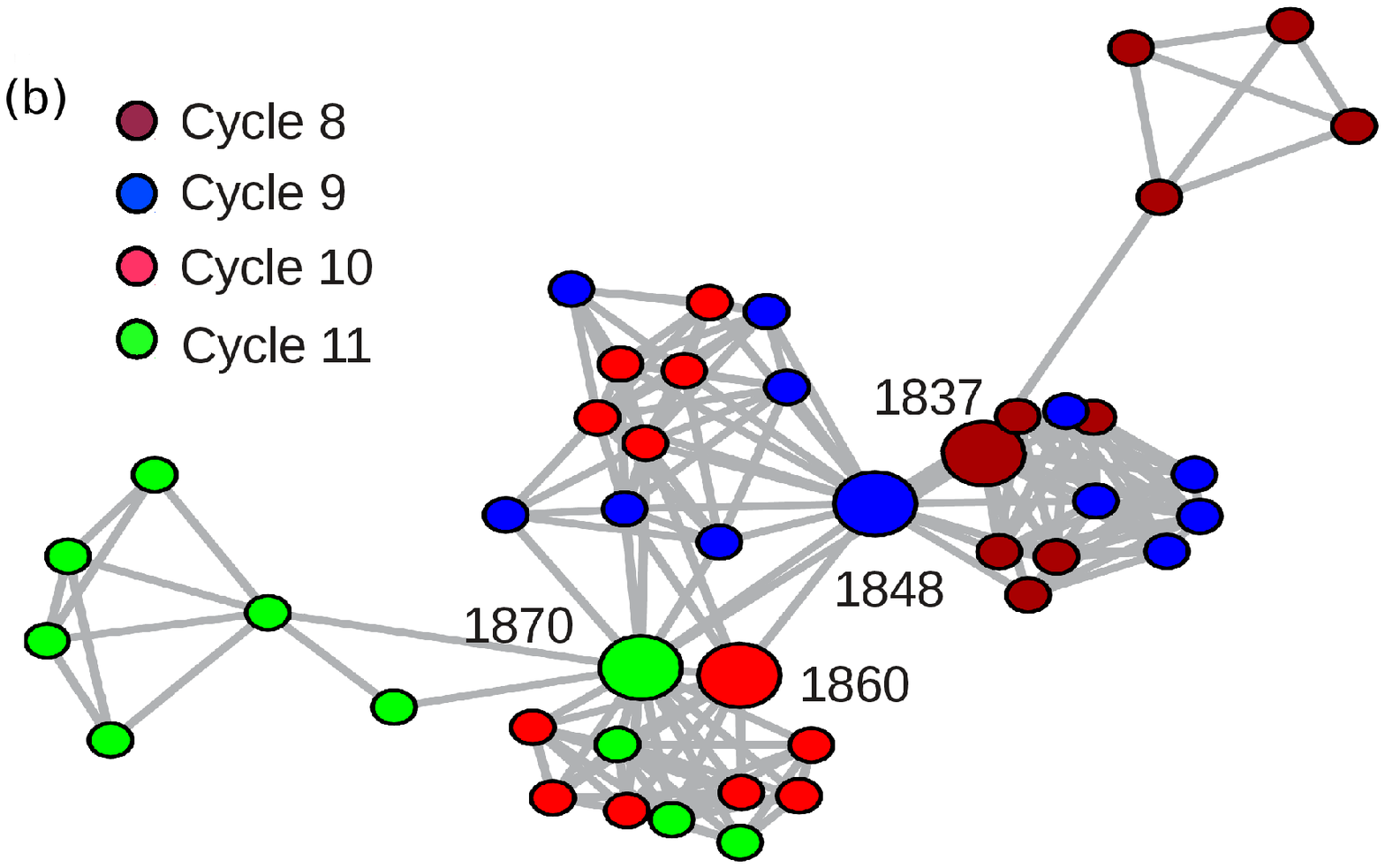}
   \includegraphics[width=\columnwidth]{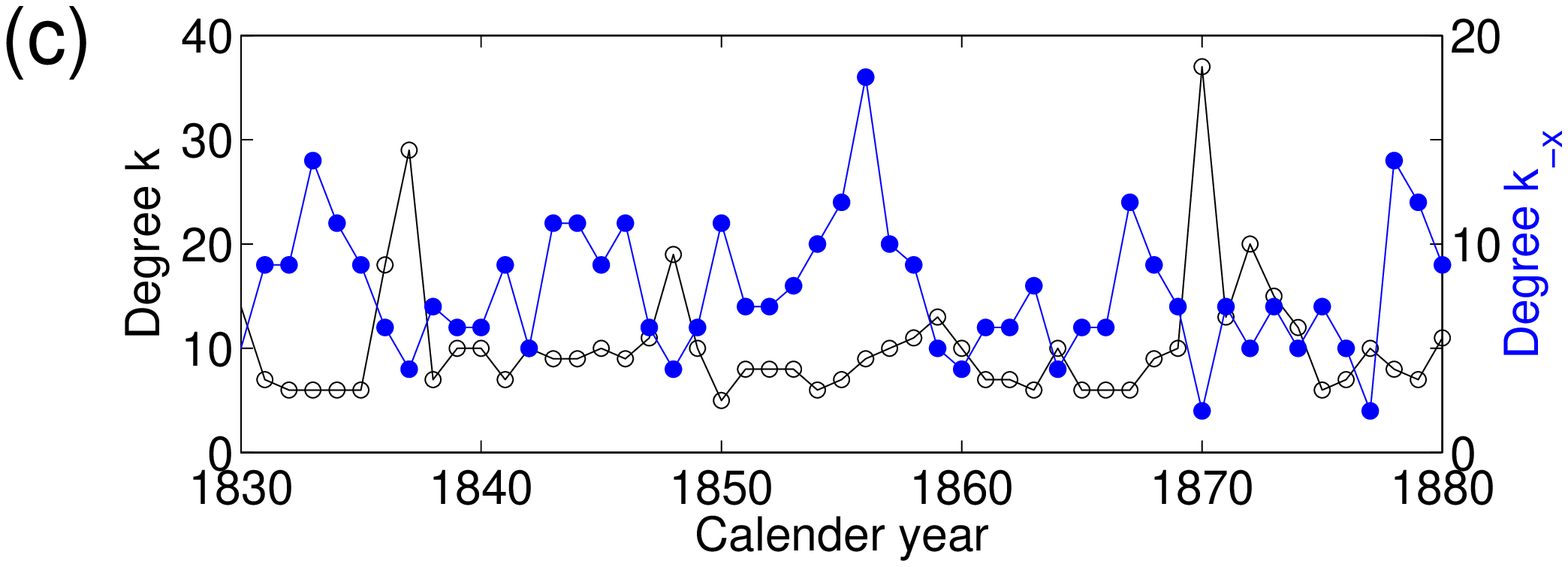}
\caption{The procedure to construct a VG. Note that  an enlargement of the
particular time span of the annual data is used here only for visualization
purposes, while the analysis performed throughout is based on both monthly and
daily series. (A) sunspot numbers in gray bar plot, where two time points
fulfilling the visibility condition are connected by a line; (B) complex network
representation, one network cluster usually includes time points of two
subsequent solar cycles; (C) degree sequences of the original series $x(t_i)$
(open circles) and the negatively inverted series $-x(t_i)$ (filled circles),
respectively.}
\label{visi_intro}
\end{figure}

Figure~\ref{visi_intro} illustrates an example of how we construct a VG for the
sunspot time series. It is well known that the solar cycle has an approximately
11-year period, which shows that most of the temporal points of the decreasing
phase of one solar cycle are connected to those points of the increasing phase
of the next cycle (Figure \ref{visi_intro}(a)). Therefore, the network is
clustered into communities, each of which mainly consists of the temporal information of
two subsequent solar cycles (Figure \ref{visi_intro}(b)). When
the sunspot number reaches a  stronger but more infrequent extreme  maximum, we have
inter-community connections, since they have a better visibility contact with
more neighbors than other time points -- hence,  forming hubs in the graph. The
inter-community connections extend over several consecutive solar cycles
encompassing the temporal cycle-to-cycle information.

Depending on various notions of ``importance'' of a vertex with respect to the
entire network, various centrality measures have been proposed to quantify the
structural characteristics of a network ({\emph{c.f.}} \cite{Newman2003}).
Recent work on VGs has mainly concentrated on the properties of the degree and its
probability distribution $p(k)$, where degree $k$ measures the number of direct
connections that a randomly chosen vertex $i$ has, namely, $k_i = \sum_j A_{i,j}$. The
degree sequence reflects the maximal visibility of the corresponding observation
in comparison with its neighbors in the time series (Figure~\ref{visi_intro}(c)).
Based on the variation of the degree sequence $k_i$, we consider $1837$, $1848$,
$1860$, and $1870$ as hubs of the network, which can be used to identify the
approximately 11-year cycle reasonably well (Figure~\ref{visi_intro}(b)).

Furthermore in the case of sunspot time series, one is often required to
investigate what contributions local minimum values make to the network --
something that has been largely overlooked by the traditional VGs. One simple
solution is to study the negatively inverted counterpart of the original time
series, namely, $-x(t_i)$, which quantifies the properties of the local minima.
We use $k_{-x}$ and $p(k_{-x})$ to denote the case of $-x(t_i)$.
Here, we remark that this simple inversion of the time series
allows us to create an entirely different complex network -- this is because
the VG algorithm itself is extremely simple and does not attempt to reconstruct
an underlying dynamical system. As shown in Figure~\ref{visi_intro}(c), $k_{-x}$
captures the variation of the local minima rather well. We will use this
technique later to understand the long-term behavior of strong minima of the
solar cycles.

The degree distribution $p(k)$ is defined to be the fraction of nodes in the
network with degree $k$. Thus if there are $N$ nodes in total in a network and
$n_k$ of them have degree $k$, we have $p(k) = n_k / N$. For many networks from
various origins, $p(k)$ has been observed to follow a power-law behavior: $p(k)
\sim k^{-\gamma}$. In the case of VGs, $p(k)$ is related to the dynamical
properties of the underlying processes~\cite{Lacasa2008}. More specifically, for
a periodic signal, $p(k)$ consists of several distinct degrees indicating the
regular structure of the series; for white noise, $p(k)$ has an exponential
form; for fractal processes, the resulting VGs often have power law
distributions $p(k) \sim k^{-\gamma}$ with the exponent $\gamma$ being related
to the Hurst exponent $H$ of the underlying time series~\cite{Lacasa2009}. It is
worth pointing out that when one seeks to estimate the exponent $\gamma$ it is
often better to employ the cumulative probability distribution $F(k) =
\sum_{k>k_0} p(k)$ so as to have a more robust statistical fit.

Estimating the exponent $\gamma$ of the hypothetical power law model for the
degree sequence of VG can be done rather straightforwardly, but, the statistical
uncertainties resulting from the observability of sunspots are a challenge for reliable
interpretation. Meanwhile, fitting a power law to empirical data and assessing
the accuracy of the exponent $\gamma$ is a complicated issue. In general, there
are many examples which had been claimed to have power laws but turn out not to
be statistically justifiable. We apply the recipe of~\inlinecite{Clauset2009} to
analyze the hypothetical power-law distributed data, namely, (i) estimating the
scaling parameter $\gamma$ by the maximum likelihood (ii) generating a $p$-value
that quantifies the plausibility of the hypothesis by the
Kolmogorov\,--\,Smirnov statistic.

Aside from the aforementioned degree $k_i$ and degree distribution $p(k)$, in
the Appendix \ref{sapp:bet}, some alternative higher order network measures are
suggested which may be applied to uncover deeper dynamical properties of the
time series from the VG.

\section{Results}
\begin{figure}
  \centering
	\includegraphics[width=\columnwidth]{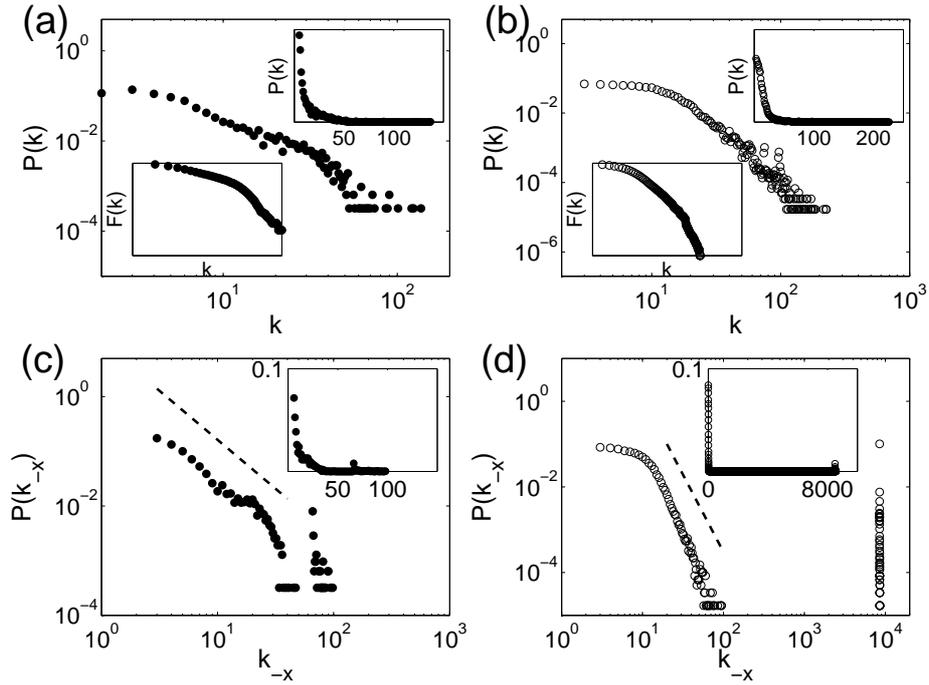}
\caption{Degree distribution $p(k)$ of VGs from monthly (A,C) and
daily data (B,D). (A,B) is for $x(t_i)$, and (C,D) $-x(t_i)$. Upper
insets of all plots are $p(k)$ in linear scale, while lower inset of (A,B) shows
cumulative distribution $F(k)$ in double logarithmic scale, where a straight
line is expected if $F(k)$ would follow a power law $\sim k^{-(\gamma - 1)}$. In
(A,B) $\gamma$s could be suspected to be in the range $[2.32,2.64]$, while a fit
to the first part of $p(k_{-x})$ yields that the slope of dashed line in (C) is
1.79, and that of (D) is 3.61, {\emph{but}} all $p$-values are $0$, rejecting
the hypothetical power laws.} \label{sn_sa_data}
\end{figure}

Figure~\ref{sn_sa_data}(a,b) show the degree distributions $p(k)$ of the VGs
derived from the ISN $x(t_i)$ with heavy-tails corresponding to hubs of the
graph, which clearly deviates from Gaussian properties.
In contrast, $p(k_{-x})$ of the negatively inverted sunspot series $-x(t_i)$
shows a completely different distribution, consisting of a bimodal property
(Figure~\ref{sn_sa_data}c,d), extra large degrees are at least two orders of
magnitude larger than most of the vertices (Figure~\ref{sn_sa_data}(d)).

\begin{figure}
	\centering
\includegraphics[width=\columnwidth]{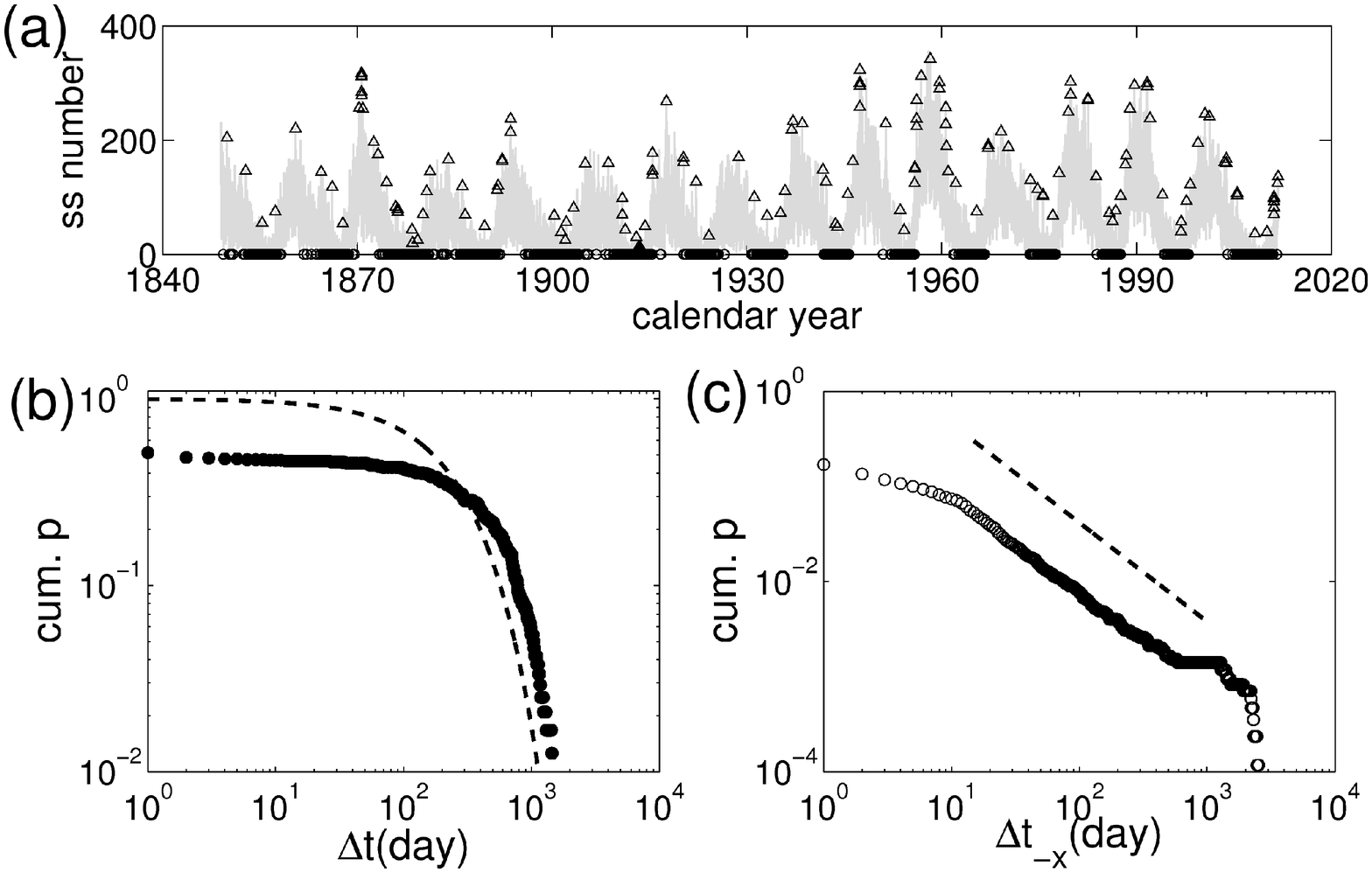}
\caption{(A) Time points annotated by vertices with large degrees.
Triangles (strong maxima) correspond to the VG constructed from the original series
$x(t_i)$, while circles (strong minima) are that of the VG from $-x(t_i)$. (B,
C) Cumulative probability distribution of the time intervals between subsequent
strong maxima (B), and minima (C). A cumulative exponential distribution is
plotted as a dashed line in (B), while the dashed line in (C) is a linear fit
with slope being equal to $\gamma - 1 = 1.04$. The corresponding $p$-value of
(C) is $0.45$, indicating that the power law is a plausible hypothesis for the
waiting times of strong minima. }
\label{ts_deg_maxMin_cp}
\end{figure}
Since well-defined scaling regimes are absent in either $p(k)$ or $p(k_{-x})$
(nor do they appear in the cumulative distributions as shown in the insets, see
captions of Figure~\ref{sn_sa_data} for details of the statistical tests which we apply), we may
reject the hypothetical power laws -- in contrast to
what has been reported in other contexts~\cite{Lacasa2008,Lacasa2009}.

\begin{figure}
  \centering
	\includegraphics[width=\columnwidth]{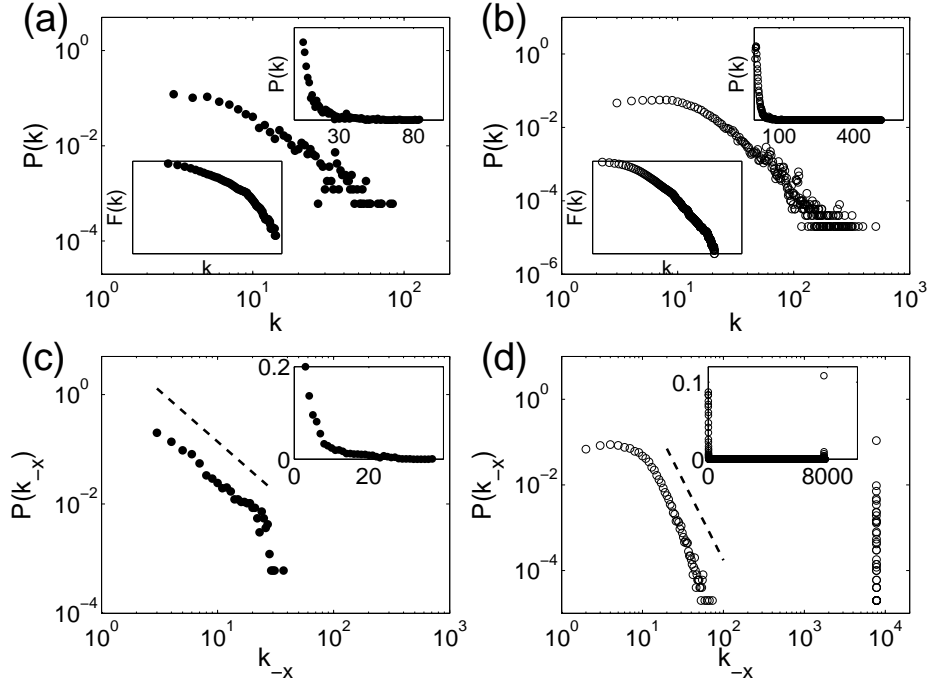}
\caption{
For the SSA data. Degree distribution $p(k)$ of VGs from monthly (A,C) and
daily data (B,D). (A,B) is for $x(t_i)$, and (C,D) $-x(t_i)$. Upper
insets of all plots are $p(k)$ in linear scale, while lower inset of (A,B) shows
cumulative distribution $F(k)$ in double logarithmic scale, where a straight
line is expected if $F(k)$ would follow a power law $\sim k^{-(\gamma - 1)}$. 
In (C,D), $\gamma$s could be fit by dashed lines (slopes: C, $1.86$; D, $3.73$
respectively), {\emph{but}} all $p$-values are $0$ rejecting the hypothetical
power laws.} \label{sa_nasa_data}
\end{figure}

\begin{figure}
	\centering
	\includegraphics[width=\columnwidth]{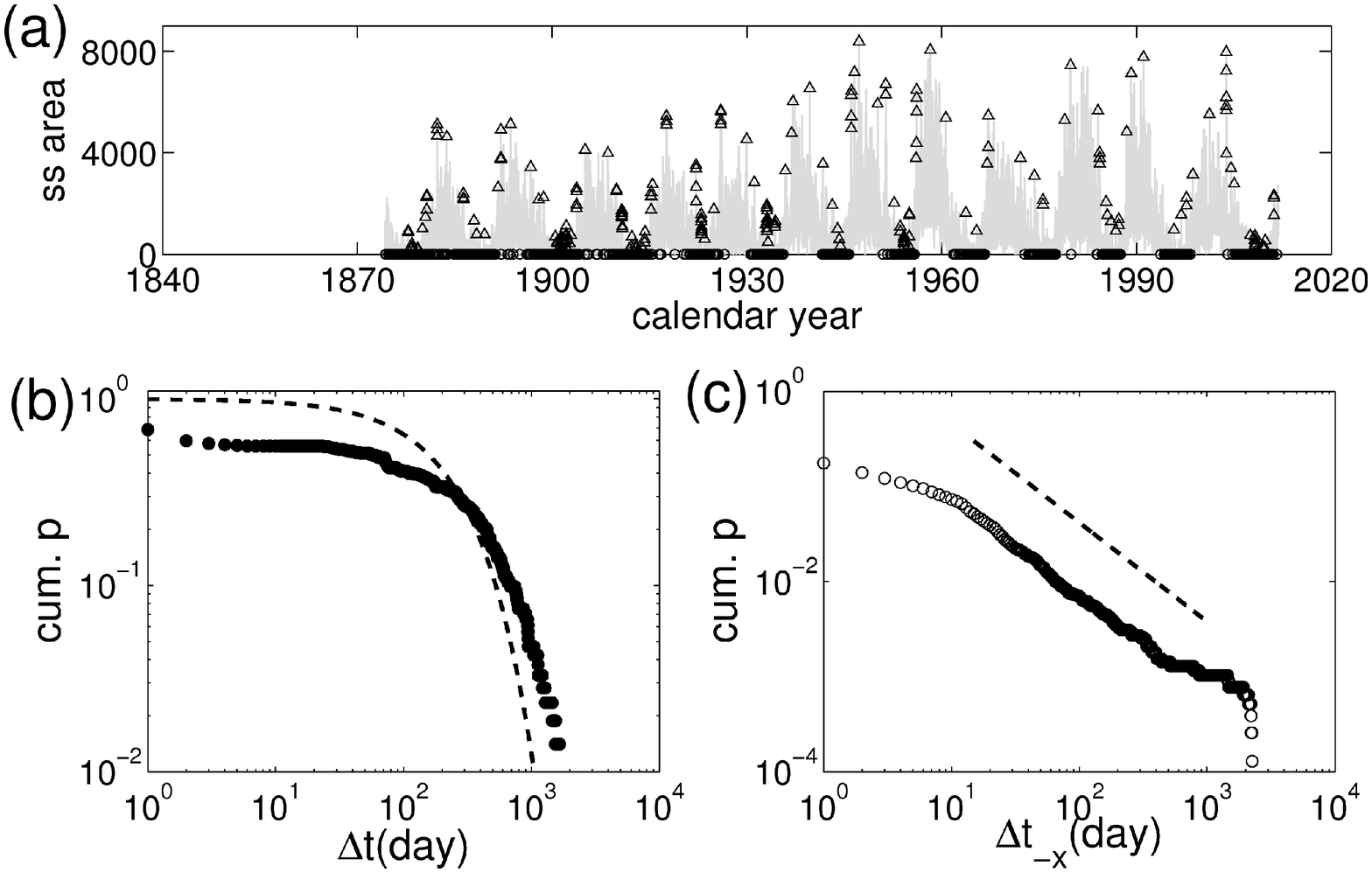}
\caption{For the SSA data. (A) Time points annotated by vertices with large degrees.
Triangles (strong maxima) correspond to the VG constructed from the original series
$x(t_i)$, while circles (strong minima) are that of the VG from $-x(t_i)$. (B,
C) Cumulative probability distribution of the time intervals between subsequent
strong maxima (B), and minima (C). A cumulative exponential distribution is
plotted as a dashed line in (B), while the dashed line in (C) is a linear fit
with slope being equal to $\gamma - 1 = 1.04$. The corresponding $p$-value of
(C) is $0.78$, indicating that the power law is a plausible hypothesis for the
waiting times of strong minima.} \label{ts_deg_maxMin_cpNASA}
\end{figure}

\begin{figure}
	\centering
\includegraphics[width=\columnwidth]{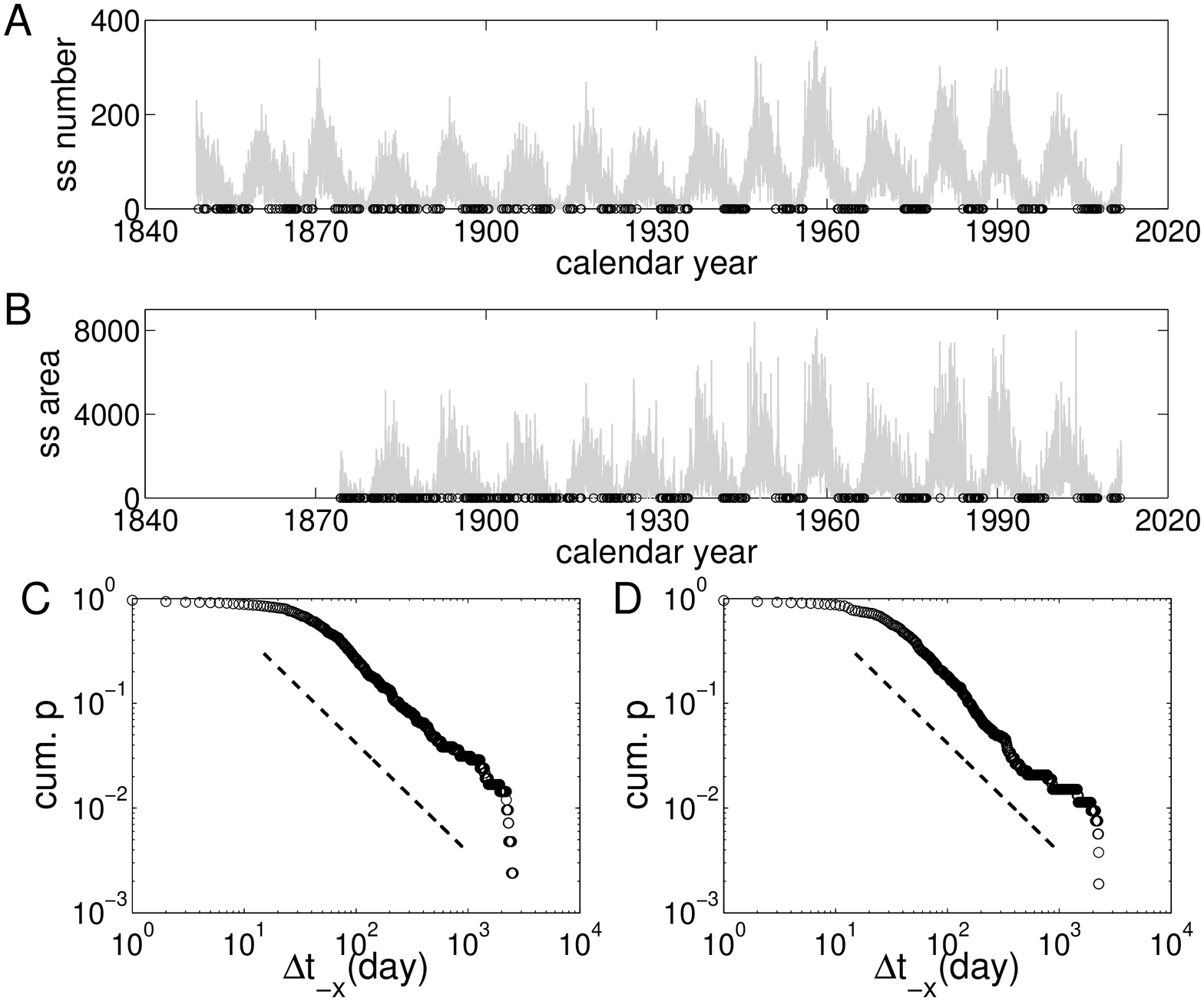}
\caption{Time points annotated by vertices with strong minima, which are defined by
network degrees ($k > 7000$) of VGs reconstructed from negatively inversed series. (A) ISN, and (B) SSA. Cumulative probability
distribution of the time intervals between subsequent strong minima, (C) ISN,
and (D) SSA. The dashed lines in (C, D) are linear fits which are obtained in
the same way as shown in Figs.~\ref{ts_deg_maxMin_cp}C,
\ref{ts_deg_maxMin_cpNASA}C, respectively. }
\label{ts_degK7000}
\end{figure}

In the context of studying grand maxima/minima over (multi-)millennial timescale
using some particular indirect proxy time series, the main idea lies in an
appropriately chosen threshold, excursions above which are defined as grand
maxima, respectively, below which are defined as grand minima
\cite{Voss1996,Usoskin2007AA}. In this work, we use a similar concept but here
in terms of degrees of the corresponding VG. We define a strong maximum if its
degree $k$ is larger than $100$ (Figure~\ref{sn_sa_data}a,b), a strong minimum
if its degree $k_{-x}$ is over $1000$ (Figure \ref{sn_sa_data}c,d). Note that,
in general, our definition of strong maxima/minima coincides \emph{neither} with
the local maximal/minimal sunspot profiles since degree $k$ takes into account
the longer term inter-cycle variations \emph{nor} with those defined for an
individual cycle. Our definition avoids the choice of maximum/minimum for one
cycle, which suffers from moving-average effects
({\emph{e.g.}},~\cite{Hathaway2010}).
In contrast, our results below are robust with respect to the
choice of the threshold degrees, especially in the case of the definition of
strong minima, since large degrees are very well separated from others. The
gray line in Figure~\ref{ts_deg_maxMin_cp}(a) shows the sunspot numbers overlaid
by the maxima/minima identified by the large degrees. We find that the positions of
strong maxima are substantially homogeneously distributed over the time domain,
while that of the strong minima are much more clustered in the time axis
although irregularly (Figure~\ref{ts_deg_maxMin_cp}(a)). We
emphasize that the clustering behavior of the strong minima on the time axis as
shown in Figure~\ref{ts_deg_maxMin_cp}(a) and
Figure~\ref{ts_deg_maxMin_cpNASA}(a) does {\emph {not}} change if threshold
degrees are varied in the interval of (200, 8000) to define strong minima,
{\emph{i.e.}}, $k > 7000$, as shown in Figure \ref{ts_degK7000}(a,b).
Therefore, the bimodality as observed in Figure~\ref{sn_sa_data}(c,d) and Figure~\ref{sa_nasa_data}(c,d) are not due to
the finite size effects of time series.

The hidden regularity of the time positions of maxima/minima can be further
characterized by the waiting time distribution~\cite{Usoskin2007AA}: the
interval between two successive events is called the waiting-time. The
statistical distribution of waiting-time intervals reflects the nature of a
process which produces the studied events. For instance, an exponential distribution is an
indicator of a random memoryless process, where the behavior of a system does
not depend on its preceding states on both short or long time scales. Any
significant deviation from an exponential law suggests that the underlying event
occurrence process has a certain level of temporal dependency. One
representative of the large class of non-exponential distributions is the power
laws, which have been observed in many different contexts, ranging from the
energy accumulation and release property of earthquakes to social contacting
patterns of humans~\cite{WuPNAS2010}.

In the framework of VGs, the possible long temporal correlations are captured by
edges that connect different communities (the increasing and decreasing phases
of one solar cycle belong to two temporal consecutive clusters). As shown in
Figure~\ref{ts_deg_maxMin_cp}(b), the distribution of the waiting time between
two subsequent maxima sunspot deviates significantly from an exponential function,
although the tail part could be an indicator of an exponential form. In
contrast, we show in Figure~\ref{ts_deg_maxMin_cp}(c) that the waiting times
between subsequent strong minima have a heavy-tail distribution where the exponent
$\gamma \approx 2.04$ is estimated in the scaling regime. This suggests that the
process of the strong minima has a positive long term correlation, which might
be well developed over the time between 15\,--\,1000 days where a power-law fit
is taken. Waiting time intervals outside this range are due to either noise effects
on shorter scales or the finite length of observations on longer time scales.
Again, the power law regimes identified by the waiting time
distributions are robust for various threshold degree values in the interval of 
(200, 8000)  (for instance, the case of $k > 7000$ is shown in
Figure \ref{ts_degK7000}(c,d)). 


\section{Discussion}
In contrast to the computations described above with
observational data, we now demonstrate the inadequacy of two models of solar
cycles. By applying the VG methods to model simulations we demonstrate that the
observed data has, according to the complex network perspective, features absent
in the models. We first choose a rather simple yet stochastic model which
describes the temporal complexity of the problem, the Barnes
model~\cite{BarnesModel1980}, consisting of an autoregressive moving average
ARMA(2, 2) model with a nonlinear transformation
\begin{eqnarray}
 z_i & = & \alpha_1 z_{i-1} + \alpha_2 z_{i-2} + a_i - \beta_1 a_{i-1} - \beta_2
 a_{i-2},  \\
 s_i &=  & z_{i}^2 + \gamma(z_i^2 - z_{i-1}^2)^2, 
\end{eqnarray} 
where $\alpha_1 = 1.90693$, $\alpha_2 = -0.98751$, $\beta_1 = 0.78512$,
$\beta_2=-0.40662$, $\gamma = 0.03$ and $a_i$ are identically independent
distributed Gaussian random variables with zero mean, and standard deviation SD
= 0.4. The second model is a stochastic relaxation Van der Pol oscillator which
is obtained from a spatial truncation of the dynamo equations~\cite{MininniPRL2000}. The equations read
\begin{eqnarray}
\dot{x} & = & y, \\
\dot{y} & = & -\omega^2 x - \mu y [3(\xi_0 + r \xi_{s})x^2-1],
\end{eqnarray}
where $\omega = 0.2993$, $\mu = 0.2044$, $\xi = 0.0102$, $\xi_{s}$ is Gaussian
noise with zero mean and SD = 1, and $r$ is adjustable but often chosen to be $0.02$.
The variable $x$ is associated with the mean toroidal magnetic field, and
therefore the sunspot number is considered to be proportional to $x^2$, which
prompts us to construct VGs from $x^2$ ($-x^2$ respectively). Both models
reproduce the rapid growth in the increasing phase and slow decay in the
decreasing phase of the activity cycles adequately. For the truncated model of
the dynamo equations, a statistical significant correlation between
instantaneous amplitude and frequency has been established, while the Barnes
model shows virtually no correlation~\cite{MininniSoPh2002}, which are generally
termed as the Waldmeier effect.

\begin{figure}
	\centering
\includegraphics[width=\columnwidth]{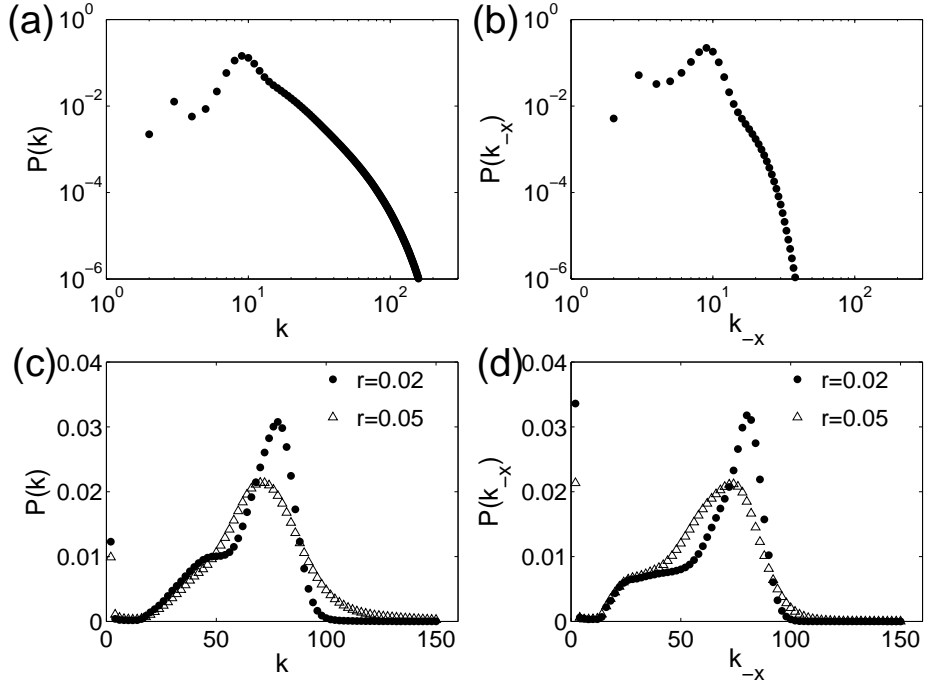}
\caption{Degree distribution $p(k)$ of VGs constructed from:
(A,~B) $x(t_i)$ and $-x(t_i)$ of Barnes' model, and (C,~D) $x^2$ and $-x^2$ of
Mininni's model. All $p(k)$ and $p(k_{-x})$ are estimated by an average over
$10,000$ independent realizations using a kernel smoother. In (C,~D) $r
= 0.02$ (filled circles), $r = 0.05$ (open triangles).}
\label{barnes_mininni_model}
\end{figure}

From both models, we generate $10,000$ independent realizations, each of them
has a one month temporal resolution and the same time span as we have for the
observations (namely, over $300$ years). We then construct VGs from both $x_{t_i}$ and
$-x_{t_i}$ from each realization in the same way as we processed for the
original observation. As shown in Figure~\ref{barnes_mininni_model}, neither
$p(k)$ can mimic the heavy tails of the distributions we observed in
Figure~\ref{sn_sa_data}(a,b), nor $p(k_{-x})$ can capture the bimodality of the
large degrees for strong minima as we have observed in the case of observational
raw records in Figure~\ref{sn_sa_data}(c,d).
This does not occur even if the parameter $r$ of the second model is adjusted
(Figure~\ref{barnes_mininni_model}(c,d)). One reason for the
absence of the bimodality of $p(k)$ in the nonlinear-oscillator model is the
fact that the model was designed to reproduce smoothed sunspot-number time
series. The often-used 13-month running-average method in the literature is
known to suppress the maximum/minimum amplitudes of the
series~\cite{Hathaway2010}. Therefore, the visibility condition for each time
point is changed if the 13-month smoothing technique is applied to the original
data. We show the degree distribution $p(k)$ of VGs
reconstructed from smoothed ISN series in Appendix (Figure~\ref{pdf_smooth}(b)),
where the bimodality is absent. 

\begin{figure}
  \centering
   \includegraphics[width=\columnwidth]{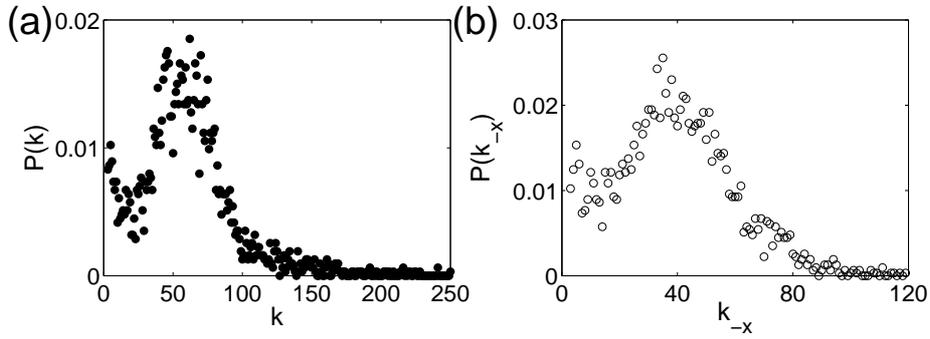}
\caption{Degree distribution $p(k)$ of VGs reconstructed from smoothed monthly ISN data. (A)
is for $x(t_i)$, and (B) $-x(t_i)$. }
\label{pdf_smooth}
\end{figure}

As shown in Figure \ref{barnes_mininni_model}, strong maxima/minima are
{\emph{not}} well separated. Consequently we are prevented from identifying unique
waiting time sequences. The corresponding analysis then depends significantly on the
choice of threshold degrees.

Dynamo theory provides several hints that might explain the features observed in
the long-term evolution of the solar activity. The two theoretical models tested
in this work can reproduce qualitative features of the system reasonably well,
however, within the context performed in this study, cannot yield a complete
and conclusive rendition of the statistical properties of $p(k)$. The power-law
regimes obtained from waiting time sequences suggest that the interaction
patterns for two subsequent minima can be much more complicated than what has
been previously described as the instantaneous amplitudes\,--\,frequencies
correlation using rather simple models~\cite{PalusPRL1999,MininniSoPh2002}.

\section{Conclusions}
In this work, we apply a recently proposed network approach, namely the
visibility graph, to disclose the intricate dynamical behavior of both strong
maximal and minimal sunspot numbers with observational records. More
specifically, we show that:
\begin{enumerate}
\item There is a strong degree of memory af the time scale of 15\,--\,1000 days
in the occurrence of low-activity episodes, observed as clusters of inactive
days. The identified persistence time scale of the strong minima agrees with the
recently proposed hypothetical long range memory on time scales shorter than the
11-year cycle~\cite{Rypdal_JGR2012}.

\item The occurrence of high activity episodes is nearly random, {\emph{i.e.}}
strong active regions appear more or less independently of each other. The
distinctive long-term correlations of the strong maxima and strong minima are
reflected by the structural asymmetries of the degree distributions $p(k)$ of
the respective VGs.

\item There is no evidence for a long term inter-cycle memory. This is in
agreement with the present paradigm based on alternative methods (see,
{\emph{e.g.}}, reviews by~\cite{Petrovay2010,Usoskin2008}), and provides an
observational constraint for solar-activity models.

\end{enumerate}
Since the long term intra-cycle memory is relatively easy to establish but
inter-cycle memory remains largely unclear, therefore we propose that our
results could be used for evaluating models for solar activity at this time
scale because they reflect important properties that are not included in other
measures reported in the literature.

From the methodological perspective, we propose an interesting generalization
for the construction of VGs from the negatively-inverted time series. This has
been seen, via our analysis of sunspot observations, to show complementary
aspects of the original series. Note that the negatively-inverse transformation
is crucial for understanding when asymmetry is preserved in the time series.
Therefore, it is worth analyzing the dependence of the resulted VGs on an
arbitrary monotonic nonlinear transformation. Furthermore, as presented in
Appendix~\ref{sapp:cor}, a systematic investigation of the general conclusion as
to whether large sampled points correspond to hubs of VGs will be a subject of
future work\,--\,especially in the presence of cyclicity and asymmetry.

It is worth stressing that we construct VGs directly based on the raw sunspot
series without any preprocessing. Many researchers prefer to base their studies
on some kind of transformed series since most common methods of data analysis in
the literature rely on the assumption that the solar activity follows Gaussian
statistics~\cite{PesnellReview2012}. It is certain that the conclusions will
then show some deviations depending on the parameters chosen for the
preprocessing. The pronounced peaked and asymmetrical sunspot-cycle profiles
prompt one to develop techniques such that the possible bias due to the
unavoidable choice of parameters should be minimized. The complex network
perspective offered by VG analysis has the clear advantage of being independent
of any priori parameter selection. 

The procedure for network analysis outlined here can be directly applied to
other solar activity indicators, for instance, the total solar irradiance and
the solar\,--\,flare index. We compared our results to two rather empirical
models and showed that the distinctive correlation patterns of maximal and
minimal sunspots are currently absent from these two models. Using our analysis
for more refined dynamo-based models ({\emph{e.g.}}~\cite{Charbonneau2010})
would be straightforward. Certainly, further work on this line of research will
examine any differences given by the particular quantity and strengthen the
understanding of the hypothetical long-range memory process of the solar
activity from a much broader overview.

\begin{acknowledgements}
This work was partially supported by the German BMBF (projects PROGRESS), the
National Natural Science Foundation of China (Grant No. 11135001), and the Hong
Kong Polytechnic University Postdoctoral Fellowship. MS is supported by an
Australian Research Council Future Fellowship (FT110100896).
\end{acknowledgements}

\appendix

\section{Graph Visualization of Degree $k_i$ and Betweenness Centrality
$b_i$}\label{sapp:bet} 
Besides the fact that the maximal sunspot numbers are identified as hubs of VGs
by the degree sequence $k_i$, convincing links between further network-theoretic
measures and distinct dynamic properties can provide some additional interesting
understanding for the time series~\cite{Donner2012}. In this study, we provide
a graphical visualization on the relationship between degree $k_i$ and node
betweenness centrality $b_i$, which characterizes the node's ability to
transport information from one place to another along the shortest path.

\begin{figure}
  \centering
   \includegraphics[width=\columnwidth]{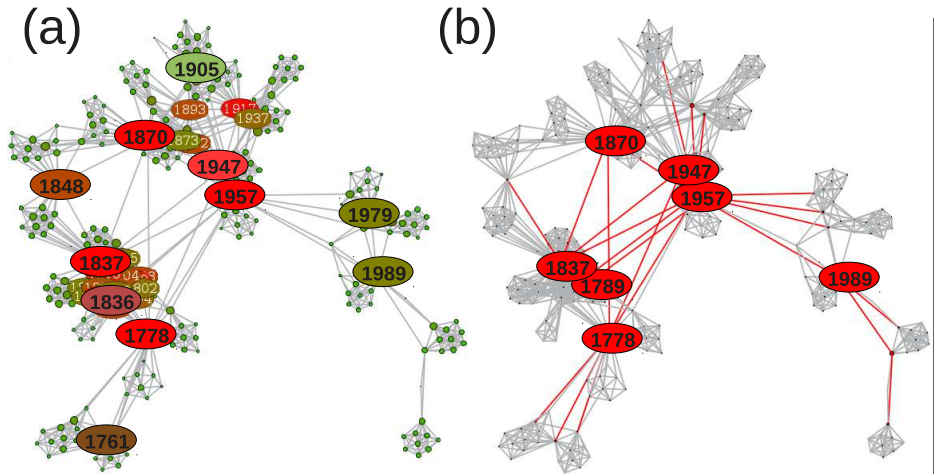}
\caption{Network representations of the VG constructed from the
annual sunspot numbers of the entire series. Highlighted visible nodes are:
(A) large degrees ($k_i>15$), and (B) high betweenness centrality ($b_i>0.2$). }
\label{year_sspnDegBet}
\end{figure}

Using the annual ISN as a graphical illustration, here only the relationship
between some relatively large degrees ($k_i > 15$) and betweenness centrality
values ($b_i > 0.2$) is highlighted in Figure~\ref{year_sspnDegBet} for the
entire series available. For instance in the annual series, time points $1778$,
$1789$, $1837$, $1870$, $1947$, $1957$, and $1989$ are all identified
simultaneously as large degrees and high betweenness, indicating strong positive
correlations.

Certainly many other measures can be directly applied to the sunspot numbers,
however, providing the appropriate (quantitative) interpretations of the results
in terms of the particular underlying geophysical mechanisms remains a
challenging task and is largely open for future work. Note that there is in
general a strong interdependence between these different network structural
quantities.

\section{Correlation between $k_i$ and $x_i$}\label{sapp:cor}
A general rule of understanding the scale-free property of the degree
distribution of complex networks is the effects of a very few hubs having a
large amount of connections. In the particular case of VGs, hubs are related to
maxima of the time series since they have better visibility contact with more
vertices than other sampled points. However, this result can not be generalized
to all situations, for instance, it is easy to generate a time series  such that its maxima 
are not always mapped to hubs in the VGs.

\begin{figure}
	\centering
	\includegraphics[width=\columnwidth]{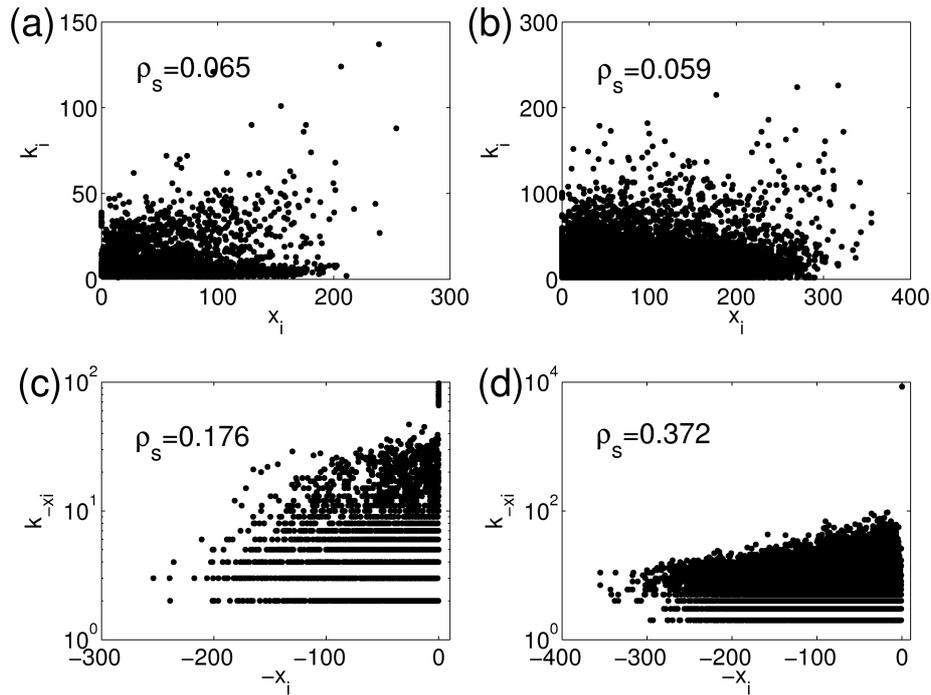}
\caption{Scatter plot of the degree sequence $k_i$\,--\,$x_i$ for VG
constructed from ISN series based on (A) monthly and (B) daily data
respectively. Spearman $\rho$ is indicated. (C, D) are based on the VGs
constructed from negatively inverted series $-x_i$. }
\label{deg_ts_rhoB}
\end{figure}

\begin{figure}
	\centering
	\includegraphics[width=\columnwidth]{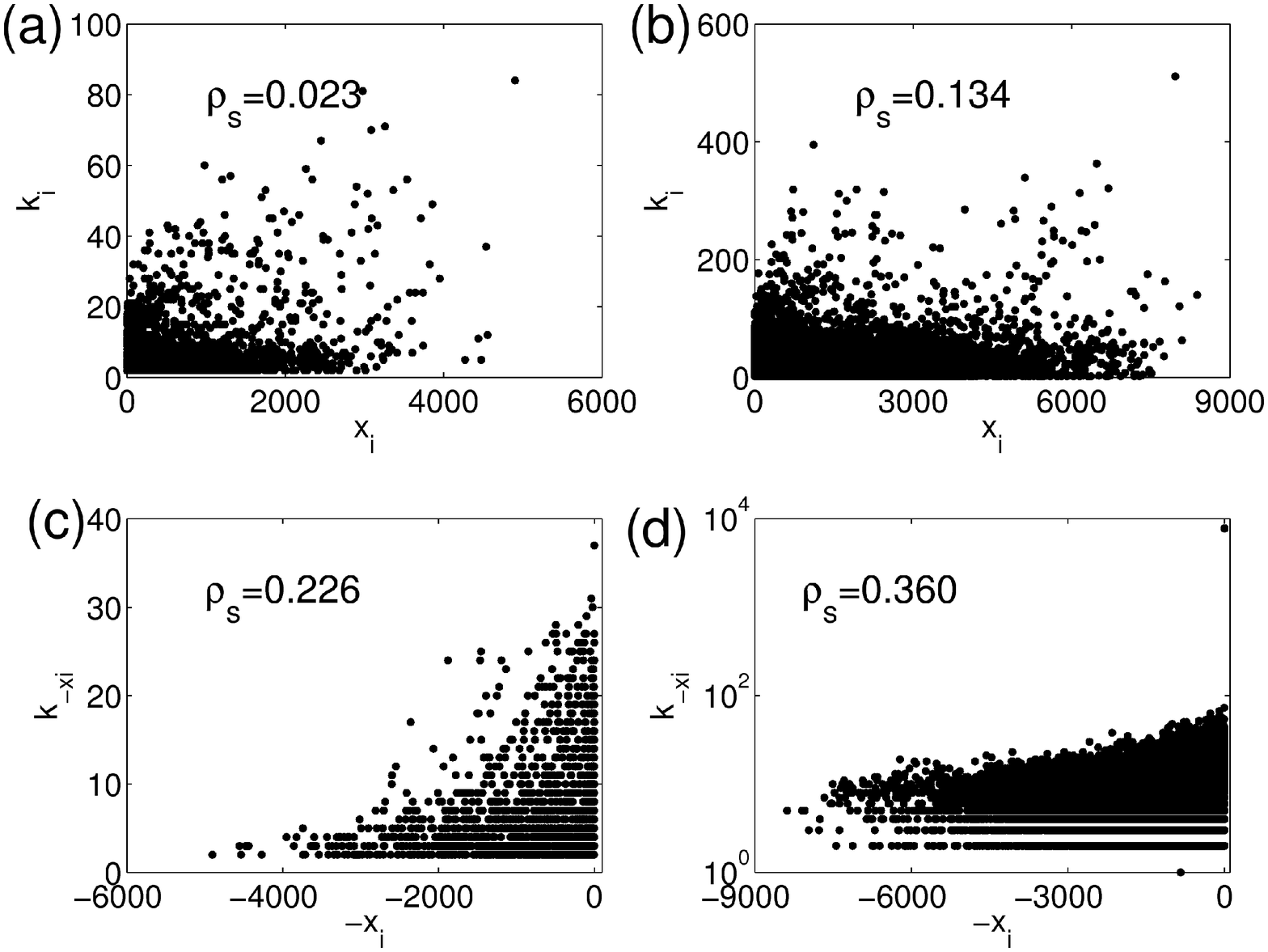}
\caption{Scatter plot of the degree sequence $k_i$\,--\,$x_i$ for VG
constructed from SSA series based on (A) monthly and (B) daily data
respectively. Spearman $\rho$ is indicated. (C, D) are based on the VGs
constructed from negatively inverted series $-x_i$.} \label{deg_ts_rhoG}
\end{figure}

One simple way to better explore this correlation is  to use scatter plots  between the
degree sequence and the sunspot time series. As we show in
Figure~\ref{deg_ts_rhoB}, the Spearman correlation coefficients $\rho$ as very
small (still significantly larger than zero) in the case of VGs reconstructed
from the original time series. This provides an important  cautionary note on the interpretation
of hubs of VGs by local maximal values of the sunspot numbers. On the contrary
if the network is reconstructed from $-x$, hubs of VGs could be better
interpreted by local minimal values of the sunspot data since the correlations
become larger. These results hold for both the ISN and SSA series
(Figure~\ref{deg_ts_rhoG}). One reason for the  lack of strong correlation between
the degree $k_i$ and $x_i$ is because of the (quasi-)cyclicity of the particular
time series, which has a similar effect as the Conway series \cite{Lacasa2008}.
It is this concave behavior over the time axis (although quasi-periodic from
cycle to cycle) that prevents the local maxima from having highly connected
vertices. It remains unclear how local maxima of a time series are mapped to
hubs of VGs -- we defer this topic for future work especially in the presence
of cyclicity. This situation becomes even more challenging if some sort of
asymmetric property is preserved in the data, as we have found for the
sunspot series.

% \bibliographystyle{spr-mp-sola}
% \bibliography{grl_ref}

\end{article} 
\end{document}